\documentclass{aastex}
\usepackage{emulateapj5}
\usepackage{graphicx}
\usepackage{times,psfig}

\def\ltsima{$\; \buildrel < \over \sim \;$}
\def\lsim{\lower.5ex\hbox{\ltsima}}
\def\gtsima{$\; \buildrel > \over \sim \;$}
\def\gsim{\lower.5ex\hbox{\gtsima}}

\newcommand{\be}{\begin{equation}}
\newcommand{\en}{\end{equation}}

\def\cmdue {\rm \ cm^{-2}}

\begin{document}

\title{A metal rich molecular cloud surrounds GRB 050904 at redshift 6.3}
\author{
S. Campana\altaffilmark{1},
D. Lazzati\altaffilmark{2},
E. Ripamonti\altaffilmark{3},
R. Perna\altaffilmark{4},
S. Covino\altaffilmark{1},
G. Tagliaferri\altaffilmark{1},
A. Moretti\altaffilmark{1},
P. Romano\altaffilmark{1},
G. Cusumano\altaffilmark{1},
G. Chincarini\altaffilmark{1,5}
}
\altaffiltext{1}{INAF-Osservatorio Astronomico di Brera, Via Bianchi 46,
I--23807 Merate (Lc), Italy}
\altaffiltext{2}{JILA, Campus Box 440, University of Colorado, 
Boulder, CO 80309-0440, USA}
\altaffiltext{3}{Kapteyn Astronomical Institute, 
University of Groningen, Postbus
800, 9700AV Groningen, The Netherlands}
\altaffiltext{4}{INAF - Istituto di Astrofisica spaziale e Fisica 
Cosmica, Via La Malfa 153, I-90146 Palermo, Italy}
\altaffiltext{5}{Universit\`a degli studi di Milano-Bicocca,
Dipartimento di Fisica, Piazza delle Scienze 3, I-20126 Milano, Italy}


\begin{abstract}
GRB~050904 is the gamma-ray burst with the highest measured
redshift. We performed time resolved X-ray spectroscopy of the late
GRB and early afterglow emission. We find robust evidence for a
decrease with time of the soft X-ray absorbing column. We model the
evolution of the column density due to the flash ionization of the GRB
and early afterglow photons. This allows us to constrain the
metallicity and geometry of the absorbing cloud. We conclude that the
progenitor of GRB~050904 was a massive star embedded in a dense metal
enriched molecular cloud with $Z\gsim0.03\,Z_\odot$. This is the first
local measurement of metallicity in the close environment of a GRB and
one of the highest redshift metallicity measurements. We also find that the
dust associated with the cloud cannot be similar to that of our Galaxy
but must be either sizably depleted or dominated by silicate grains.
We discuss the implications of these results for GRB progenitors and
high redshift star formation.  
\end{abstract}

\section{Introduction}
The studies of the properties of the interstellar medium (ISM) at high
redshift have traditionally relied on the observations of quasars and
most of our knowledge about the ISM metallicity comes from
measurements of the absorption signatures of overdense gas along their
lines of sight.  However, such measurements are likely giving us a
partial view of the properties of the high redshift medium, as they
are biased towards regions of low density (e.g. Bromm \& Larson 2004).
On the other hand, quasar spectra also show emission lines coming from
dense gas in the immediate vicinity of the quasars themselves.  They
provide strong hints of an high metal content: for example,
millimetric and radio observations of a CO line from the most distant
quasar ($z=6.4$), suggest the presence of large amounts of carbon and
oxygen (Walter et al. 2003; Bertoldi et al. 2003). The main drawback of
such measurements is that they necessarily probe only the high
luminosity end of the quasar luminosity function, whereas it is known
that at low redshifts the measured metallicity scales with luminosity
(Pettini et al. 1997; Pentericci et al. 2002).

A fundamental complement to studies performed with quasars in order to
have a more representative and comprehensive understanding of the
properties of the ISM in high-redshift galaxies comes from gamma--ray
burst (GRBs) studies (Fiore et al. 2005, Prochaska, Chen \& Bloom 2006).  
GRBs are intense, short
impulses of $\gamma$--rays occurring randomly in the sky. Their
intense luminosity allows their detection up to the highest redshifts
(in principle $z\gsim 20$), outshining by orders of magnitude the one
from quasars at similar distances.  Long GRBs are associated with the
death of massive stars (Galama et al. 1998; Stanek et al. 2003; Hjorth
et al. 2003; Malesani et al. 2004; Campana et al. 2006) and should occur 
in star forming
regions.  This property, together with their power law, almost
featureless spectra, make GRBs ideal probes of star forming regions in
the high-redshift Universe. What particularly differentiates GRB
sources from quasars as cosmological lighthouses is the fact that,
being very bright and very short-lived, GRBs can alter the equilibrium
state of their surrounding medium on an observable timescale by
photoionizing gas (Perna \& Loeb 1998) and destroying dust (Waxman \&
Draine 2000). These effects cause a gradual reduction of the opacity
from the X--ray band (due to progressive ionization of the metals) to
the optical band (due to the progressive destruction of the dust
grains).

Here we use time dependent X-ray spectroscopy to investigate the
properties of the environment of GRB~050904 (Cusumano et al. 2006;
Kawai et al. 2006), the farthest detected so far at a redshift of
$z=6.3$ (Tagliaferri et al. 2005; Kawai et al. 2006). We show that the
burst most likely exploded within a rather compact HII region,
produced inside a dense molecular cloud during the main sequence life
of the progenitor star. We also show that a substantial metal
enrichment must have taken place before the formation of the
progenitor star, pointing at the presence of a generation of massive
stars at $z>6$, even in objects of relatively low luminosity and mass
like the GRB host galaxy (Berger et al. 2006).

\section{Swift data}

The Swift satellite (Gehrels et al. 2004) discovered GRB 050904
with its Burst Alert Telescope and re-pointed its X--ray and
UV/optical telescopes within 161~s (Cusumano et
al. 2006). Observations continued up to ten days.  Data were
accumulated in Window Timing (WT) up to 534 s after the trigger and in
Photon Counting (PC) later on (see Burrows et al. 2005a). Data were
extracted using the latest software version (HEAsoft 6.0.4) and a
standard ($40\times 20$ pixel) window for WT data and a circular (30
pixel radius) region for PC data. A small hole in the extraction
region of 2 pixels has been used when the PC data showed signs of
pile-up.

The number of photons collected by the Swift X--Ray Telescope (XRT)
was sufficiently high to allow us to perform time resolved spectral
analysis of the X--ray light curve of GRB~050904 (Cusumano et
al. 2006; Watson et al. 2006; Gendre et al. 2006). Analysis was 
carried out in the 0.3--10 keV for WT and PC modes. Spectra are
grouped to 30 counts per spectral bin. A systematic error of $3\%$ has
been added to cope with the uncertainties in the XRT response
matrices. To describe the GRB spectra when flares are present, we adopted
a power law emission model with a high energy cut-off (Burrows et
al. 2005b; Falcone et al. 2006). We divided the XRT light curve into four 
time intervals and computed the equivalent column density with the above 
spectral model. Given the large number of X--ray flares (see Fig. 1 in Cusumano
et al. 2006) we fit the spectra (three in WT and one in PC\footnote{XRT
response matrices for the two observing modes have been cross-calibrated on a stable
X--ray sources without showing any major difference.}) with
different cut-off energies and power law spectral indices. The
contribution of our Galaxy to the absorption toward GRB~050904 can be
estimated from HI maps as $4.9\times10^{20}\cmdue$.  IRAS maps instead
provide a lower value of $3.5\times10^{20}\cmdue$.  We left the Galactic column 
density free to vary in the range
$(3-5)\times10^{20}\cmdue$ (using the model {\tt tbabs}, within
XSPEC). The absorption in the host galaxy was modelled with the
{\tt zvfeabs} within XSPEC, setting the the iron abundance to zero
since at such an early time it had likely not been produced yet
(e.g. Prochaska et al. 2002; allowing for the presence of iron would
result in an inferred column density about $20\%$ smaller). Finally,
we fit the spectra minimising the gain shift in order to overcome
small energy scale problems, especially relevant at low
energies\footnote{see
http://swift.gsfc.nasa.gov/docs/heasarc/caldb/swift/docs/xrt/SWIFT-XRT-CALDB-09.pdf
a more complete description will be presented in Campana et al. (2007, in
preparation).}. 

We considered three possible metal abundances (0.1, 0.3 and 1
$Z_\odot$ except of iron, see also below). We detect highly significant 
absorption in excess of the Galactic value. For all the considered 
metallicities we find that the equivalent column density decreases with 
time at a high significance level (see Table 1, see also Watson et al. 2006; 
B\"oer et al. 2006; Gendre et al. 2006, Fig.~\ref{fig:fit}). In
particular, we note that the GRB spectrum gets softer as time passes (see
Table 1). If the column density decrease would be a spurious effect one would
expect an increase in the observed column density, being larger the number of
photons at low energies to be hidden. This is not the case, strenghtening the observational
result. This is consistent with the idea that the circum-burst absorbing
material is photoionized by the high-energy photons of the burst. Among the
very few GRBs for which opacity evolution has been detected, this is not
only the first at a known distance, but also the one at the largest
distance known so far. This is not surprising since the cosmological
time dilation allows us to observe the X--ray spectrum at small
comoving times.

\begin{table*}
\label{data}
\caption{Results of the spectral fit with variable absorption and metallicity.}
{\small
\begin{tabular}{ccc|ccc|c}
\hline
Time interval & XRT obs.& Total && $N_H$ &($10^{22}$ cm$^{-2}$) \\
(s rest frame)& mode$^*$ & counts& ($Z=0.1\,Z_\odot$)    &($Z=0.3\,Z_\odot$) &
($Z=Z_\odot$) & Power law ($\Gamma$)   \\
\hline
$28.9\pm6.9$     &WT&3674&$71.3^{+14.1}_{-14.1}$ 
                 &$25.1^{+3.9}_{-4.6}$ 
                 &$7.9^{+1.1}_{-2.1}$ & $1.2^{+0.1}_{-0.1}$\\
$49.5\pm13.7$    &WT&4729&$66.5^{+7.5}_{-7.6}$ 
                 &$23.3^{+3.0}_{-4.8}$ 
                 &$7.3^{+1.4}_{-1.4}$ & $1.5^{+0.1}_{-0.1}$\\
$73.5\pm10.3$    &WT&1532&$41.0^{+9.7}_{-10.3}$ 
                 &$14.4^{+4.3}_{-3.8}$ 
                 &$4.5^{+2.3}_{-1.1}$ & $1.8^{+0.1}_{-0.3}$\\
$4130.5\pm4051.8$&PC&4987&$15.4^{+3.7}_{-5.5}$ 
                 &$5.3^{+2.1}_{-1.8}$ 
                 &$1.8^{+0.7}_{-0.4}$  & $1.6^{+0.2}_{-0.2}$\\
\hline
Probability$^+$& &       & $1.6\times 10^{-10}$ &$1.4\times 10^{-7}$ &$3.2\times 10^{-6}$\\
Reduced $\chi^2$ (nhp)$^\dag$ & &&1.00 ($48\%$)& 1.00 ($48\%$)     & 1.00 ($48\%$)\\ 
\hline
\end{tabular}
} 

Errors are at $1\,\sigma$ confidence level ($\Delta \chi^2=1.0$).  

$^+$ This probability has been estimated as the probability that a
constant $N_H$ model provides an acceptable fit to the data.

$^\dag$ We have 304 degrees of freedom in each fit. In parenthesis are reported
the null hypothesis probabilities (nhp) for each model.

\end{table*}

The data quality does not allows us to fit the metallicity
as a free parameter. An upper limit on the initial column density of
the material along the line of sight comes from the consideration that
if it were larger than $1/\sigma_T=1.5\times10^{24}$~cm$^{-2}$, the
medium would be Thomson thick and the prompt $\gamma$-ray radiation
multiply scattered. This would introduce a variable delay between
photons, attenuate the prompt emission, and smear the variability
pattern, in contrast with the observations. By imposing a Thomson
thin medium at the time of the measurements, we obtain the constraint
on the metallicity $Z\gsim 0.03\,Z_\odot$ ($90\%$ confidence level,
see Fig. 1).

\

\section{Photoionization study}

The propagation of the photons through the cloud surrounding the GRB
alters the local thermal equilibrium conditions. The gas ionization
state and dust content will then depend on the time from the burst
onset and the distance from the burster. The optical depth of the
medium, which is obtained by radial integration of the local opacity
at a given time, can be computed only numerically (Perna \& Lazzati
2002).  We use here a time-dependent photoionization and dust
destruction code that can take into account any light curve,
metallicity and dust content (Perna \& Lazzati 2002). It allows us to
simulate how the conditions in the environment of GRB~050904 vary as a
result of the burst radiation. Inputs of the code are the total
hydrogen column density (irrespective of its ionization state) and the
geometry of the absorber, that we assume to be a shell at distance $R$
from the burst site and width $\Delta R=0.1R$. The choice of fixing
the shell width is due to the consideration that the density of the
absorber is irrelevant, due to the fact that the recombination times
are much longer than the observed times.  The fit results are reported
in Table~2. We find that the absorption and its evolution can be
explained by a GRB surrounded by a shell of material with a radius of
a few parsecs and high column density $N_H\sim
7\times10^{22}/(Z/Z_\odot)$ (see Fig.~\ref{fig:chi}). This is
consistent with X-ray spectral fitting data (see Fig. 1) and it
implies a total mass of the absorber of
$M\sim{\rm{few}}\times10^5\,M_\odot$ and a total mass of metals
(excluding He) of about $10^{2\div3}\,M_\odot$. Higher metallicities
imply larger radii and therefore larger metal masses. In all cases,
the fits have an acceptable value of $\chi^2$ (associated probability
larger than $\sim 50\%$). We also explored the role of the initial gas
temperature. We find that for initial temperatures $T_0<10^4$~K the
fits are indistinguishable. If the initial temperature is instead above
$10^4$~K, HII starts to become relevant and the best fit radius moves
outwards. The column density is instead unaffected.

\begin{table*}
\label{tab:fit}
\caption{Result of the time dependent column density fits.}
{\small
\begin{tabular}{cccccc}
\hline
$Z/Z_\odot$ & $T(0)^*$ (K)    & $N_H^\dag$  ($10^{22}$ cm$^{-2}$) & $R$ (pc)
& Mass ($10^4\,M_\odot$) & $\chi^2$ (prob.)$^+$   \\
\hline
0.1         & $10^4$        & $64\pm12$               & $1.2\pm0.3$ & $9^{+3.5}_{-1.5}$  & 1.4 ($50\%$)   \\
0.3         & $10^4$        & $23^{+9}_{-5}$          & $1.8\pm0.7$ & $9\pm5$   & 1.2 ($55\%$) \\
1.0         & $10^4$        & $7.7\pm1.5$             & $3.2\pm0.6$ & $12\pm3$  & 0.86 ($65\%$)   \\
1.0         & $1.5\times10^4$ & $7.7\pm1.1$           & $4.7\pm0.9$ & $24\pm6$  & 0.79 ($67\%$)   \\
\hline
\end{tabular}
}

Errors are at $1\,\sigma$ confidence level for one parameter of interest ($\Delta \chi^2=1.0$). 

$^*$ Initial temperature of the gas before the GRB explosion.

$^\dag$ $N_H$ is the column density of hydrogen irrespective of ionization $N_H=N_{HI}+N_{HII}$.

$^+$ $\chi^2$ for the fits with 2 degrees of freedom. In parenthesis
are reported the null hypothesis probabilities. Given the quality of the fit,
all the fit are statistically acceptable.

\end{table*}

These results allow us to infer the origin of the absorbing medium. It
must be located in the close vicinity of the GRB progenitor, and so it
could be either the molecular cloud where the progenitor star formed
or the material ejected from the stellar progenitor during its
evolution, in the form of a wind or of a massive ejection
event. However, the mass of the absorber is too large to be the result
of ejection from a single stellar progenitor (not even from a massive
Population III star). We conclude that the progenitor of GRB~050904
formed within a molecular cloud enriched with metals. 
In support of our interpretation, Frail et al. (2006) found a density of 
$\sim700$~cm$^{-3}$ at a distance of $\sim 1$~pc from the burster from the
modelling of the radio afterglow, while Kawai et al. (2006) and Totani et
al. (2006) find a density of $n\sim 200$~cm$^{-3}$ at a distance larger than
several tens of parsecs from fine structure lines in the SiII absorption.
In Fig.~\ref{fig:chi} we overlay the 
locus of HII regions produced by massive stars on the probability
best-fit contours. We find that the properties of the best fit region
resemble those of an HII region, especially in conditions where most
of the H is ionized ($T_0=1.5\times10^4$~K). The high temperature
simulation also shows a residual HI column that could explain the late
time observation of $N_{HI}\sim4\times10^{21}$~cm$^{-2}$ in the
optical spectrum several days after the explosion (Totani et al. 2006)
in the case of a molecular cloud with roughly solar composition.

\begin{figure}
\centerline{\includegraphics[width=\columnwidth]{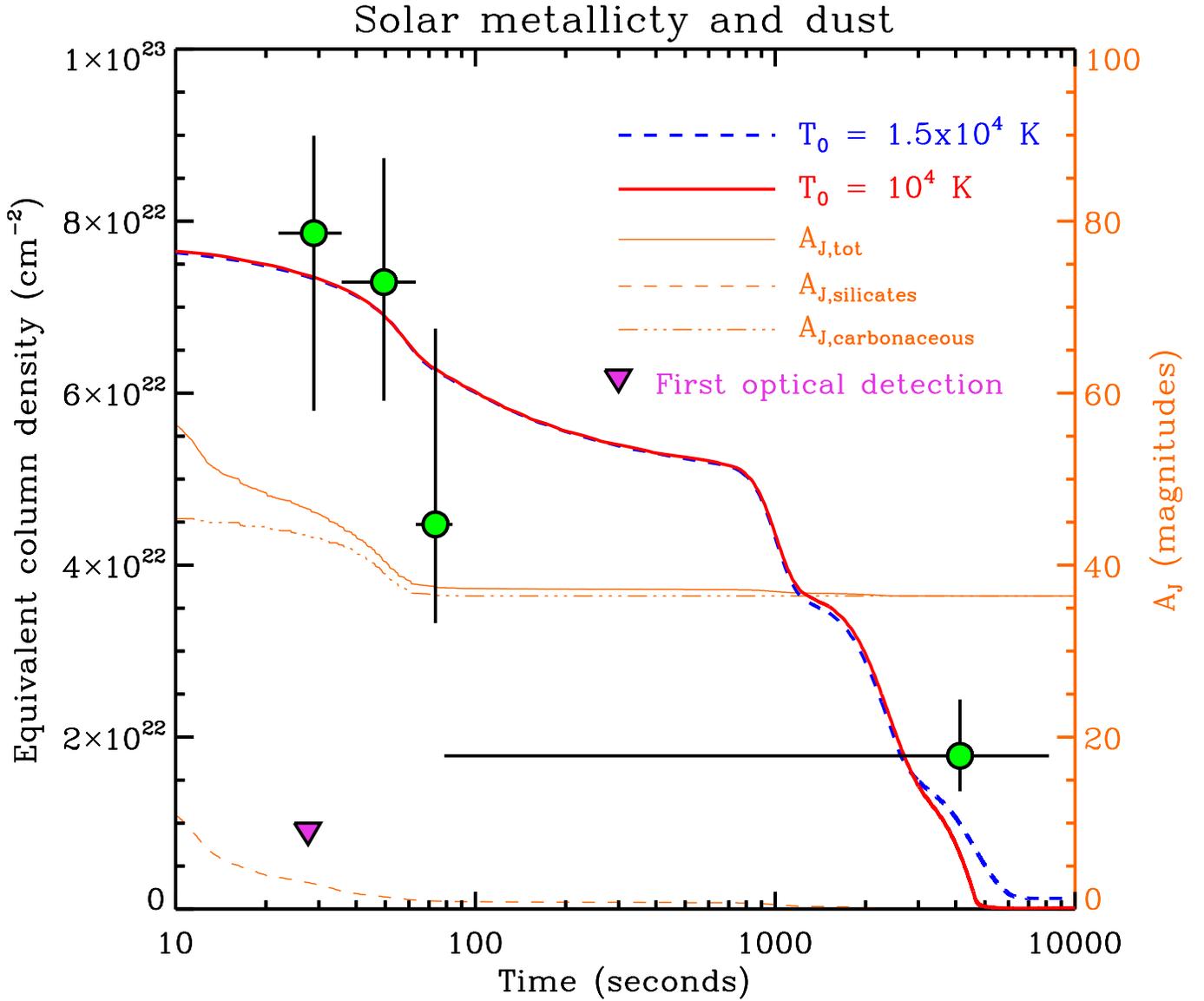}}
\caption{Evolution of the equivalent column density measured in the
X--ray afterglow of GRB~050904 (green dots with errors at
$1\,\sigma$).  Time is in the rest frame. Solar metallicity with no Fe
and Ni has been assumed. The equivalent column density is defined as
the column density that would produce the same amount of absorption
for a cold non-ionized absorber.  The red solid and blue dashed lines
show the best fit models (see Table~2) for different initial
temperatures. The photoionization code has in input the observed light
curve of GRB~050904 (Cusumano et al. 2006). The drop in absorption at
$t\sim1000$ s (in the rest frame) corresponds to the group of bright
X--ray flares.  The thin orange lines (and right y axis) show the
amount of absorption that would be observed in the J band (rest frame
$\sim 7.2$~eV) if the X--ray absorbing medium would be polluted with
Galactic-like dust (Mathis et al. 1977). The optical transient was
observed at $t_{\rm{obs}}=200$~s (i.e. 27 s rest frame, B\"oer et al. 2006) 
in white light,
indicating very little absorption. Thin orange dashed and dot-dashed
lines show the absorption due to silicates only and to carbonaceous
grains only, respectively. The little extinction implied by the early
optical observation can be explained by a dust component rich in
silicates and depleted in carbonaceous grains. This could be the
results of an ISM enriched by pair instability SNe (Schneider et
al. 2004).\label{fig:fit}}
\end{figure}

\begin{figure}
\centerline{\includegraphics[width=\columnwidth]{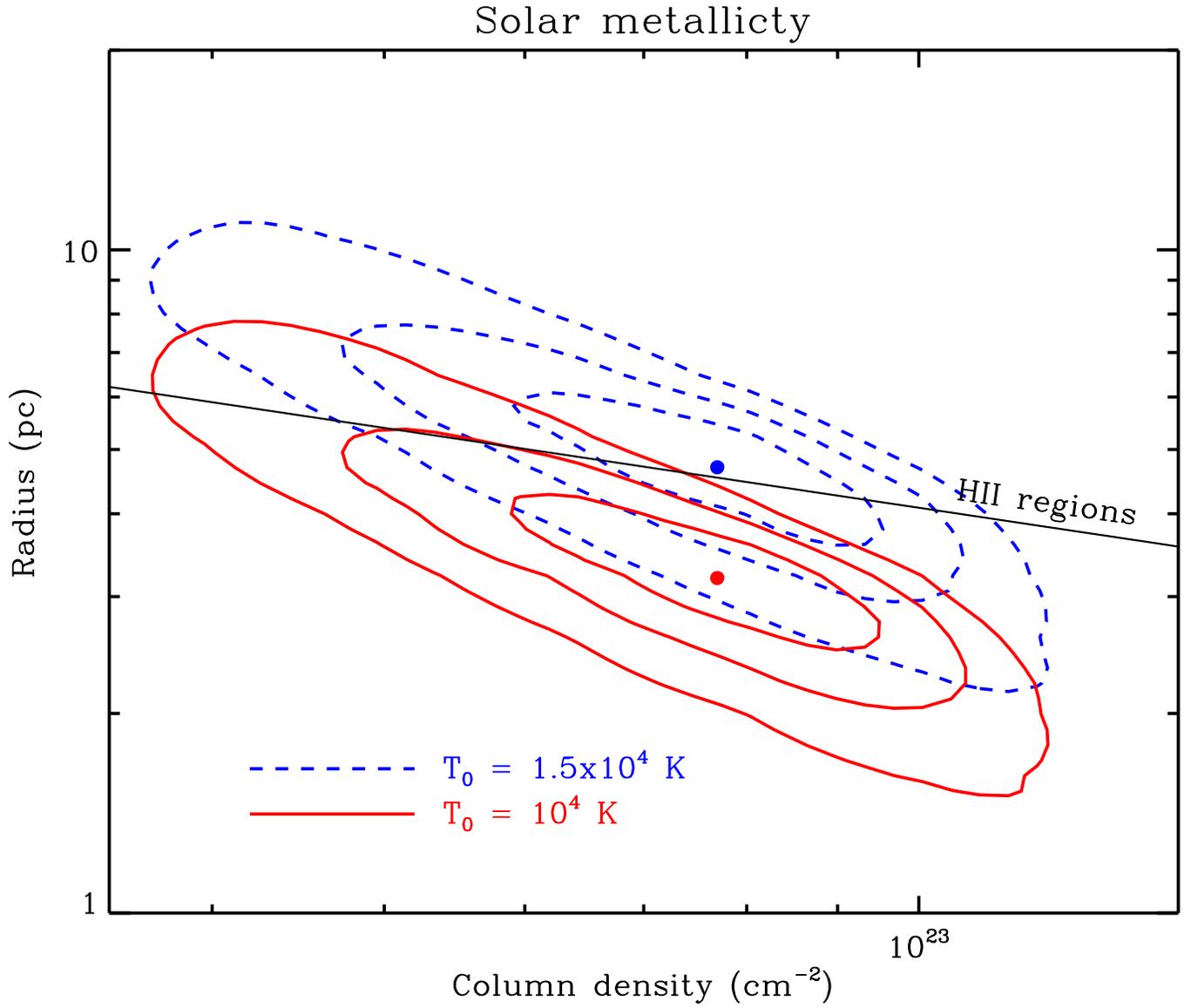}}
\caption{Confidence contour (1, 2 and 3-$\sigma$) in the radius --
column density plane for the solar metallicity case. The computed grid
of radii and column densities over which the fit was performed is much
larger, and spans $10^{16}<R<10^{21}$~cm and
$10^{21}<N_H<10^{25}$~cm$^{-2}$. The black line shows the locus of HII
regions surrounding massive stars at the end of their life under the
assumption of uniform density of the progenitor molecular
cloud. Despite the simple model, the agreement is
satisfactory. \label{fig:chi}}
\end{figure}

Optical-to-near infrared observations detected the afterglow,
resulting in a rest frame observation in the ultraviolet (Tagliaferri
et al. 2005; Haislip et al. 2006; B\"oer et al. 2006). Photometric
redshift codes were successful in recovering the spectroscopic
redshift without the addition of local reddening. In order to detect
this optical emission we have to require the almost complete absence
of dust absorption. Given the large amount of matter along the line of
sight this is not straightforward. A Galactic-like dust component
(Mathis et al. 1977) is only partially destroyed by the GRB high
energy photons, resulting in an absorption of $A_J\sim 46$ magnitudes
at the time of the first optical observation (see
Fig.~\ref{fig:fit}). The residual absorption is almost entirely due to
carbonaceous grains (see the dot-dashed line in
Fig.~\ref{fig:fit}). The lack of absorption can be therefore explained
if the GRB environment had been enriched primarily by pair-instability supernovae, 
which are expected to produce mainly silicate dust grains (Schneider et al. 2004; see 
also Todini \& Ferrara 2001).  This is not unreasonable, because the star formation rate of 
$\sim 15~M_\odot\,{\rm yr^{-1}}$ given by Berger et al. (2006) implies that 
pair-instability supernovae should be occurring at a rate of 0.01--1 per year.

\section{Conclusions}

Optical and radio observations of the afterglow of GRB~050904 showed
that it is likely embedded in a dense environment with relatively high
metallicity (Kawai et al. 2005; Berger et al. 2006). We have analyzed
the Swift-XRT X-ray data of GRB~050904. Time resolved spectroscopy of
the burst and afterglow reveals a very high column density at early
times. The absorption decreases at later stages, almost disappearing
$\sim10^4$~s after the explosion. This behavior can be understood if
the absorbing material is located very close to the GRB, so that the
flash ionization of the environment progressively strips all the
electrons off the ions decreasing the opacity of the medium. We model
the evolution of the opacity by simulating the flash ionization and
dust destruction numerically with our time dependent code (Perna \&
Lazzati 2002). We constrain the geometric properties of the
environment, finding that GRB~050904 exploded within a dense
metal-enriched molecular could. The nearby environment must have a
metallicity of at least several per cent solar. Such a metallicity is
not due the the burst progenitor but is a property of the whole
cloud. We also find that most likely the immediate surroundings of the
GRB was shaped as an HII region by the UV flux of the progenitor star.
This result is consistent with the metallicity estimate
derived from optical studies (Kawai et al. 2006; Totani et al. 2006), with a
metallicity of $\sim 0.1\,Z_\odot$ based on SII lines. This should provide a
fair description of the metallicity since sulphur is not depleted to dust
grains.

Our results confirm, albeit in an indirect way, that even the highest
redshift GRBs are associated to the death of massive stars. The
metallicity constraint, the first derived for the molecular cloud
where the burst exploded rather than for the host galaxy as a whole,
shows that GRB progenitors can have relatively high metallicity. 

From the cosmological point of view, our result shows that metal
enrichment had already taken place at redshifts larger than 6. We also
find that, for the derived abundance of metals, the absorber does not
show any evidence of dust extinction. In the Galaxy environment, a UV
extinction of several tens of magnitudes would be associated to such a
high column density. We find that the GRB radiative heating would not
have been enough to evaporate all the dust. The lack of UV extinction
can be due to a strong underabundance of dust grains with respect to
our local environment. Such an explanation seems unlikely given the
availability of metals and their SN origin (Todini \& Ferrara
2001; Sugerman et al. 2006). We find that a possible alternative explanation 
is a dust mixture biased toward silicate grains, which are more easily 
destroyed by UV heating. In this case the flash ionization of the GRB would
reduce the UV extinction to low levels. Metal and dust enrichment due
to a top heavy initial mass function (possibly dominated by pair instability 
SNe) could explain a silicate rich dust mixture (Schneider et al. 2004).

\begin{acknowledgments} 
This work is supported by ASI (I/R/039704), NASA, NSF and
NWO grants. We gratefully acknowledge the contributions of dozens of
members of the Swift team who helped make this instrument possible.
\end{acknowledgments}

\end{document}